\documentclass[preprint,aps,prd,amsmath,amssymb]{revtex4}
\pdfoutput=1
\usepackage{graphicx}
\usepackage{color}
\usepackage{graphicx}
\usepackage{dcolumn}
\usepackage{bm}
\usepackage{amsmath}
\usepackage{amsfonts}
\usepackage{bbm}
\usepackage{subfigure}
\usepackage{setspace}
\newcommand{\beq}{\begin{eqnarray}}
\newcommand{\eeq}{\end{eqnarray}}

\newcommand{\ie}{{\it i.e.}}

\newcommand{\la}{{\lambda}}

\newcommand{\morder}[1]{{\cal O}\left(#1 \right)}
\newcommand{\eq}[1]{(\ref{#1})}

\newcommand{\nn}{\nonumber}
\newcommand{\beqa}{\begin{eqnarray}}
\newcommand{\eeqa}{\end{eqnarray}}
\newcommand{\bea}{\begin{eqnarray}}
\newcommand{\eea}{\end{eqnarray}}
\newcommand{\be}{\begin{equation}}
\newcommand{\ee}{\end{equation}}
\newcommand{\beqat}{\begin{eqnarray*}}
\newcommand{\eeqat}{\end{eqnarray*}}

\newcommand{\pat}{{\partial}}

\newcommand{\inv}[1]{\frac{1}{#1}}

\newcommand{\bmp}{\noindent\begin{minipage}{16cm}}
\newcommand{\emp}{\end{minipage}\vskip 7mm} 

\def\drawbox#1#2{\hrule height#2pt
        \hbox{\vrule width#2pt height#1pt \kern#1pt
              \vrule width#2pt}
              \hrule height#2pt}

\def\Asym#1#2{\vcenter{\vbox{\drawbox{#1}{#2}
              \kern-#2pt 
              \drawbox{#1}{#2}}}}

\begin{document}
\title{\Large Holographic Conformal Window \\ ---  \\A Bottom Up Approach }
\author{Matti {\sc J\"{a}rvinen}}
\email{mjarvine@cp3.sdu.dk}
\author{Francesco {\sc Sannino}}
\email{sannino@cp3.sdu.dk}
\affiliation{CP$^{\rm \it 3}$-Origins, Campusvej 55, DK-5230 Odense M, Denmark.}

\begin{flushright}
{\it CP$^{\rm \it 3}$-Origins-2009-23}
\end{flushright}


\begin{abstract}
We propose a five-dimensional framework for modeling the background geometry associated to ordinary Yang-Mills (YM) as well as to nonsupersymmetric gauge theories possessing an infrared fixed point with fermions in various representations of the underlying gauge group. The model is based on the improved holographic approach, on the string theory side, and on the conjectured all-orders beta function for the gauge theory one. We first analyze the YM gauge theory. We then investigate the effects of adding flavors and show that, in the holographic description of the conformal window, the geometry becomes $AdS$ when approaching the ultraviolet and the infrared regimes. As the number of flavors increases within the conformal window we observe that the geometry becomes more and more of $AdS$ type over the entire energy range.  \end{abstract}

\maketitle

\section{Introduction}

Understanding strong dynamics poses a formidable challenge. For decades physicists have been working on several aspects associated to the strongly coupled regime of gauge theories of fundamental interactions such as Quantum Chromodynamics (QCD). One of the most fascinating possibilities is that strongly coupled gauge theories have magnetic dual gauge theories or higher dimensional gravity duals. A number of remarkable exact results have been found for supersymmetric gauge theories. The situation for nonsupersymmetric gauge theories still relies on a number of  speculative conjectures such as the $AdS/QCD$ one (see for a review \cite{Erdmenger:2007cm}). Nevertheless one can argue that a geometric interpretation of known gauge theory results may lead to a different way to investigate several aspects of strong dynamics. Here we  bring this speculation one step further and provide a direct investigation of the conformal window for different number of flavors via a very simple five-dimensional gravitational model. 

On the gauge theory side we have made much progress in trying to uncover the phase diagram of nonsupersymmetric gauge theories, as function of number of flavors, colors and fermionic matter representation, using analytical as well as first principle lattice methods. The knowledge of the phase diagram is relevant when constructing extensions of the standard model in which the electroweak symmetry breaks dynamically, see \cite{Sannino:2009za} for a review.

A relevant question is wether or not one can construct a dual gravitational description of the phase diagram.  One can argue that a dual description should incorporate the dependence on the number of flavors and representation with respect to the four dimensional underlying gauge theory. To achieve this goal one should have a phase diagram on the gauge theory side and a way to match it to possible gravitational duals on the other side.  The phase diagram for generic representations, gauge groups and physically relevant chiral gauge theories was pioneered in \cite{Sannino:2004qp,Dietrich:2006cm,Ryttov:2007sr,Ryttov:2007cx,Ryttov:2009yw,Sannino:2009aw,Sannino:2009za,Sannino:2009qc,Sannino:2009me}. Several recent analysis make use of further conjectures \cite{Poppitz:2009tw,Armoni:2009jn}. These methods give very close answers, within their relative uncertainties. Comprehensive reviews are available \cite{Sannino:2009za,Sannino:2008ha}. The method that best fits our present purpose is the one which makes use of a recently conjectured all orders beta function for $SU(N)$ gauge groups, put forward in \cite{Ryttov:2007cx} and generalized to any  gauge group even with chiral-like matter in \cite{Sannino:2009aw,Sannino:2009za}.  The effects of adding a mass term on the beta function appeared in \cite{Dietrich:2009ns} while a further extension trying to take into account possible extra zeros in the beta function have been investigated in \cite{Antipin:2009wr}. The original beta function \footnote{Albeit the exactness of the beta function has not yet been proven, we know of no other analytic method, used to predict the conformal window  passing the same number of consistency checks. } results are also highly consistent with the phase diagram which emerges via the exact solutions of the 't Hooft anomaly conditions investigated in \cite{Sannino:2009qc,Sannino:2009me}. Knowing a beta function on the gauge theory side immediately selects as a possible approach to investigate the gravity dual, i.e. the one envisioned in Refs.~\cite{Gursoy:2007cb,Gursoy:2007er}. Further work on this and closely related subjects has been performed in \cite{Bigazzi:2005md,Cotrone:2007gs,Gursoy:2008bu,Gursoy:2008za,Pirner:2009gr,Gursoy:2009jd,Alanen:2009ej,Gursoy:2009kk,Galow:2009kw,Alanen:2009xs}. Other approaches to holography within near conformal nonsupersymmetric theories are discussed in \cite{Carone:2006wj,Dietrich:2008ni,Dietrich:2008up,Nunez:2008wi,Mintakevich:2009wz,Elander:2009pk,Dietrich:2009af,Nunez:2009da,Alvares:2009hv}.

\section{The All orders beta function Conjecture}
To start setting the notation we recall the two loops beta function which reads:
\begin{eqnarray}\beta (g) = -\frac{\beta_0}{(4\pi)^2} g^3 - \frac{\beta_1}{(4\pi)^4} g^5 \ ,
\label{perturbative}
\end{eqnarray}
where $g$ is the gauge coupling and the beta function coefficients are given by
\begin{eqnarray}
\beta_0 &=&\frac{11}{3}C_2(G)- \frac{4}{3}T(r)N_f \\
\beta_1 &=&\frac{34}{3} C_2^2(G)
- \frac{20}{3}C_2(G)T(r) N_f  - 4C_2(r) T(r) N_f  \ .\end{eqnarray}
To this order the two coefficients are universal,
i.e. do not depend on which renormalization group scheme one has used to determine them.
The perturbative expression for the anomalous dimension to two-loops reads:
\bea
 \gamma(g) &=& a_0 \frac{g^2}{4 \pi} + a_1 \frac{g^4}{(4 \pi)^2}  + \morder{g^6}\quad \mathrm{with} \nn\\
 a_0 &=& \frac{3}{2\pi} C_2(r) \ ; \quad a_1 = \inv{16 \pi^2} \left[3 C_2(r)^2 - \frac{10}{3} C_2(r)N_f + \frac{97}{3} C_2(r)C_2(G) \right]  \ . \label{gamma}
\eea
Here $\gamma =-{d\ln m}/{d\ln \mu}$ with $m$ being the renormalized fermion mass. The generators $T_r^a,\, a=1\ldots N^2-1$ of the gauge group in the
representation $r$ are normalized according to
$\text{Tr}\left[T_r^aT_r^b \right] = T(r) \delta^{ab}$ while the
quadratic Casimir $C_2(r)$ is given by $T_r^aT_r^a = C_2(r)I$. The
trace normalization factor $T(r)$ and the quadratic Casimir are
connected via $C_2(r) d(r) = T(r) d(G)$ where $d(r)$ is the
dimension of the representation $r$. The adjoint
representation is denoted by $G$. 

The two-loop beta function above does not capture nonperturbative corrections and we now consider the beta function put forward in \cite{Ryttov:2007cx} which has the following properties: 
\begin{itemize}
\item{Reproduces the supersymmetric exact one \cite{Novikov:1983uc} when restricting to ${\cal N} = 1$ super YM.}
\item{At large $N$ reproduces the exact results via two-index symmetric/antisymmetric representation \cite{Armoni:2003gp}.}
\item{Fits the conformal window prediction via the investigated gauge-duals suggested in \cite{Sannino:2009qc}.}
\item{Provides a remarkable agreement with the running of the YM beta function obtained via lattice. More precisely the deviations from the two-loops beta function are of the same order of the lattice ones. Also the dependence purely on the 't Hooft coupling is nicely encoded and is in good agreement with data \cite{Luscher:1992zx,Luscher:1993gh,Lucini:2007sa}.}
\item{The anomalous dimensions of the mass is predicted, in a closed form, at the infrared fixed point (IRFP) \footnote{Recent lattice investigations \cite{Catterall:2007yx,Catterall:2008qk,DelDebbio:2008zf,Hietanen:2008vc,Hietanen:2009az,Pica:2009hc,Catterall:2009sb,Shamir:2008pb} of the anomalous dimensions at the fixed point \cite{Lucini:2009an,Bursa:2009we,DeGrand:2009hu,DeGrand:2008kx,DeGrand:2009mt} for (Next) to Minimal Walking Theories \cite{Sannino:2004qp,Dietrich:2005jn,Dietrich:2005wk,Foadi:2007ue}  are too preliminary to draw conclusions on their size.  Especially Next to Minimal Walking Technicolor there is no numerical convincing evidence of the presence of a fixed point yet. In the case of Minimal Walking Technicolor can happen that, the simulations are not yet able to approach the fixed point because of the presence of a lattice bulk phase transition just before reaching it. On the other hand the spectrum and several other observables might very much look conformal already. If this were the case then one is computing the anomalous dimension at a smaller value of the coupling which is not the fixed point one. It is easy to check that a small variation of the value of the fixed point coupling leads to large variations of the exact value of the anomalous dimension (see section 4.7 of \cite{Sannino:2009za}).}.}
\end{itemize}
The beta function \cite{Ryttov:2007cx} reads: 
\begin{eqnarray}\label{iBETA}
\beta(g) &=&- \frac{g^3}{(4\pi)^2} \frac{\beta_0 - \frac{2}{3}\, T(r)\,N_f \,
\gamma(g^2)}{1- \frac{g^2}{8\pi^2} C_2(G)\left( 1+ \frac{2\beta_0'}{\beta_0} \right)} \ ,
\end{eqnarray}
with
$\beta_0' = C_2(G) - T(r)N_f $.  

It is a simple matter to show that the above beta function reduces
to Eq. (\ref{perturbative}) when expanded to $O(g^5)$.  As we decrease the number of flavors from just below the point where asymptotic freedom is lost one expects a perturbative (in the coupling) zero in the beta function to occur \cite{Banks:1981nn}. From the expression proposed above one finds that at the zero:
\begin{eqnarray}
\gamma = \frac{11C_2(G)-4T(r)N_f}{2T(r)N_f} \ .\end{eqnarray}
 The value of $\gamma$ increases as we keep decreasing the number of flavors. The dimension of the chiral condensate is $D(\bar{\psi} \psi)=3-\gamma$ which at the IR fixed point value reads
\begin{equation}
D (\bar{\psi} \psi)= \frac{10T(r)N_f - 11C_2(G)}{2T(r)N_f} \ .
\end{equation}
 To avoid negative norm states in a conformal field theory one must have $D\geq 1$ for non-trivial
 spinless operators \cite{Mack:1975je,Flato:1983te,Dobrev:1985qv}. Hence the critical number of flavors below which the unitarity bound is violated is
\begin{eqnarray}
N_f^{\rm{BF}} = \frac{11}{8} \frac{C_2(G)}{T(r)} \ ,
\end{eqnarray}
which corresponds to having set $\gamma=2$. 

The analysis above summarizes the one in \cite{Sannino:2008ha} and is
similar to the one done for supersymmetric gauge theories \cite{Seiberg:1994pq}.
The actual size of the conformal window may be smaller than the one
presented here which hence can be considered as a bound on the size of the window. The reason being that chiral symmetry breaking could be triggered for a value of $\gamma$ lower than two.

 \section{Beta functions meet Holography}

G\"ursoy, Kiritsis and Nitti \cite{Gursoy:2007cb,Gursoy:2007er} recently introduced a five dimensional holographic model for QCD and YM theories. The model is partly based on a noncritical five dimensional string theory. Input from the gauge theory  side is, however, needed. In particular, the potential of the dilaton field is fixed by a beta function of QCD.

Following \cite{Gursoy:2007cb} the five dimensional action in the Einstein frame is:
\be \label{Lagr}
 \mathcal{L} = M^3 \int d^5x\sqrt{g}\left[R-\frac{4}{D-2}\frac{(\pat \la_s)^2}{\la_s^2}+V(\la_s)\right] \ .
\ee
We neglected the axion field, which can be added later. Here $D$ is the number of the space dimensions fixed to be equal to five. 
$\la_s$ is related to the dilaton $\phi$ via $ \la_s= N \exp(\phi)$ with  $N$ the number of colors.   It is expected to match with the 't Hooft coupling $\la=g^2N$ of the dual four dimensional theory in the UV. The metric is taken to be: 
\bea
 ds^2 = e^{2A(r)}\left[dr^2 + \eta_{\mu\nu} dx^\mu dx^\nu\right]  \ . \eea
$r$ is the conformal coordinate ranging from zero (in the UV) to an IR value noted by $r=r_0$ (where $r_0$ can be infinite).
In the limit $r \to 0$  one should recover the $AdS$ behavior, i.e. $A \simeq \log(\ell/r)$ with $\ell$ the $AdS$ radius. One identifies in this region the warp factor $A$ with the renormalization energy scale of QCD $\mu$ as follows: $A \sim \log \mu \sim -\log r$.

The warp factor $A$ and  $\la_s$ satisfy two equations of motion following from extremizing \eq{Lagr}. These can be manipulated in a way in which one of them is independent of the dilaton potential $V(\la_s)$, while the other depends on it.  We will fix the potential by requiring the equation of motions to yield specific solutions which are supposed to best match the gauge theory expectations as done in \cite{Gursoy:2007cb,Gursoy:2007er}. The equations of motion read:
\be \label{eom}
 \ddot A - \dot A^2 + \frac{4 \dot \la_s^2}{9 \la_s^2}=0 \ ; \qquad  \frac{\dot \la_s}{\dot A}  =  \beta_s
\ee
where dots denote derivatives with respect to $r$.

As shown in \cite{Gursoy:2007cb}, the first equation can be integrated to
\be
 \frac{dA}{dr} = - \frac{e^A}{\ell} \exp\left[-\frac{4}{9}\int_0^{\la_s}\frac{d\la \beta_s(\la)}{\la^2}\right] \ .
\ee
The second equation corresponds to the parametrization
\be
 V(\la_s) = \frac{12}{\ell^2}\left[1-\frac{\beta_s^2(\la_s)}{9 \la_s^2}\right] \exp\left[-\frac{8}{9}\int_0^{\la_s}\frac{d\la \beta_s(\la)}{\la^2}\right] 
\ee
of the dilaton potential.

The relevant field theory variables are the 't Hooft coupling $\lambda$ and the energy scale $\log \mu$. The ordinary gauge theory beta function and the {\it string} theory one are defined as:
\bea
 \beta(\lambda) =2\lambda {\frac{\beta(g)}{g}}= \frac{d\lambda}{d\log\mu} \quad {\rm and}\quad  \beta_s(\lambda_s) = \frac{d\lambda_s}{d A} \ . \label{strbe}
\eea
The following relations allow for a link between the gauge theory and the gravity dual: 
\bea
 \la = \la(\la_s) \equiv f_\la(\la_s)\ ; \qquad 
 \log \mu &=& \log \mu(A) \equiv f_A(A) \ .
\eea
The functions $f_\la$ and $f_A$ are known only in the UV regime where they approach identity. 
The quantities $\la,\mu,\la_s,$ and  $A$ are linked cyclically  as follows:
\be
  \la \ \substack{\phantom{A} \\ \longleftrightarrow \\ \beta}\  \log \mu\  \substack{\phantom{A} \\ \longleftrightarrow \\ f_\la}\  A \  \substack{\phantom{A} \\ \longleftrightarrow \\ \beta_s} \  \la_s \  \substack{\phantom{A} \\ \longleftrightarrow \\ f_A} \ \la \ .
\ee
In particular, the chain rule implies
\be \label{chrule}
 \beta(\la) = \frac{d\la}{d\log\mu} = \frac{d\la}{d\la_s} \frac{d\la_s}{d A} \frac{d A}{d \log \mu} = \frac{f_\la'(\la_s)}{f_A'(A)} \beta_s(\la_s)={\cal{F}}(\lambda_s)\beta_s(\la_s)\  .
\ee
Where in the last step, following \cite{Gursoy:2007cb,Gursoy:2007er}, we have also stated that $f_{A}^{\prime}$ is a function of $\lambda_s$. Note that in the expression above we are taking:
\begin{equation}
f_\la'(\la_s) \equiv \frac {d f_\la (\la_s)}{d\la_s} \ , \qquad f_A'(A) \equiv \frac {d f_A(A)}{dA} \ .
\end{equation}
We expect at small couplings \begin{equation} \lim_{{\lambda_s} \to 0} {\cal{F}}(\lambda_s)  = 1 \ , \end{equation}

In practice the two beta functions can be understood, from the gauge theory point of view, as the beta function in two different renormalization schemes. In \cite{Zeng:2008sx} the reader can find an earlier interesting attempt to use the all-order beta function in connection with the $AdS/QCD$.

\subsection{Constraining the IR behavior}

If both the beta functions $\beta(\la)$ and $\beta_s(\la_s)$ are known, the relation \eq{chrule} can be used to extract information on $ {\cal{F}}(\lambda_s) $. This could lead to useful hints on the $\alpha'$ corrections to $f_\la$ and $f_A$ in the IR. 

Following \cite{Gursoy:2007cb,Gursoy:2007er} one can motivate the IR asymptotic form of $\beta_s$ by requiring that one reproduces (electric) confinement, magnetic screening and linear asymptotic glueball spectrum for the YM theory, yielding:
\be \label{betaas}
 \beta_s(\la_s) = - \frac{3 \la_s}{2} \left[1+\frac{3}{8\log \la_s} + \morder{\inv{\left(\log\la_s\right)^2}}\right] \ .
\ee
In particular, it seems that the string coupling $\la_s$ must run to infinity in the IR. The asymptotic connection between $A$ and $\la_s$ is obtained by substituting \eq{betaas} into \eq{strbe}: 
\be \label{Aas}
 A(\la_s) = - \frac{2}{3} \log \la_s + \frac{1}{4}\log \log \la_s + \mathrm{const.} 
\ee

Knowing the gauge theory beta $\beta(\la)$ one can further constrain the transition functions. To achieve this goal we find convenient to rewrite \eq{chrule} using the definition of the string beta function in \eq{strbe} as
\be \label{ascond2}
 A'(\la_s) = \inv{\beta_s(\la_s)} = \frac{ f_\la'(\la_s)}{f_A'(A(\la_s)) \beta(f_\la(\la_s))}  \ .
\ee
Upon integration one finds: 
\be \label{assol}
 A(\la_s) = f_A^{-1} \left( \int^{f_\la(\la_s)}\!\!\!\frac{d\la'}{\beta(\la')}  \right) \ .
\ee

The IR {\it physical} constraints on the dependence of $A$, as function on $\la_s$, are encoded in \eq{Aas}. They appear on the left-hand side of \eq{assol}. The gauge theory information is contained in $\beta(\lambda)$. Requiring the two sides to agree allows to partially constrain  $f_A$ and $f_\la$. The latter are expected to be monotonic to ensure invertibility. In the following we will use the all-order beta function on the right-hand side of \eq{assol}.

\section{Yang - Mills}

By setting the number of flavors to zero in \eq{iBETA} the all-orders YM beta function reads: 
\begin{equation}
\beta_{YM}(g)= -\frac{g^3}{(4\pi)^2}\frac{{\beta}_0}{1-\frac{g^2}{(4\pi)^2}\frac{{\beta}_1}{{\beta}_0}}\ ,
\end{equation}
with
\begin{equation}
\beta_0=\frac{11N}{3} \ , \qquad \beta_1=\frac{34N^2}{3} \ ,
\end{equation}
respectively for the one and two loop coefficients of the beta function. It is amusing to note that the all-orders beta function, although is supposed to capture nonperturbative corrections, can be built directly via the only two universal coefficients.
This form of the beta function yields a running of the coupling  very close to the one observed via first principle lattice simulations \cite{Luscher:1992zx,Luscher:1993gh,Lucini:2007sa}.    Another property is that it only depends on the 't Hooft coupling $\la$, i.e. it does not have any explicit dependence on $N$: 
\be \label{exbeta}
 \beta_{YM}(\la) = 2\la \frac{\beta_{YM}(g)}{g} = - \frac{b_0 \la^2}{1-\la/\la_0}
\ee
 where
 \be \label{QCDval}
  b_0= \frac{11}{24\pi^2}\simeq 0.046\ ;\qquad \la_0 = \frac{88\pi^2}{17} \simeq 51 \ .
 \ee
 
 We now solve the \eq{eom} in the following three cases: i) The case in which the transition functions are the identity and the {\it string} beta function is the all-order one; ii) The case in which the transition functions are the identity and  {\it string} beta function corresponds to the universal two-loops one; iii) The case in which the IR behavior described in the previous section is reproduced with the above gauge beta function, which requires opportunely modified transition functions. We will then compare and comment the results. In the appendix we provide details on how the differential equations are solved numerically. 
 
 \subsection{$\beta_s=$ all-orders $\beta$ with $f_\la(\la_s) = \la_s$ and $f_A(A) = A$  }
 In this case one has $\la_s = \la$ and $\log\mu = A$ implying that $\beta_s (\la_s) \equiv \beta(\la_s)$ with $\beta$ here taken to be the all-orders YM beta function given above. The solutions of the \eq{eom} for the relevant quantities are presented in Fig.~\ref{fig:backgrounds}. The curves for this setup are the solid blue ones.  In the top-left panel it is plotted the evolution of the coupling as a function of the energy scale ($\log \mu \sim - \log r$).
The geometry of the solutions (top-right and bottom-left panels) has the following qualitative behavior:  The geometry is almost exactly $AdS$ but the space ends abruptly in the IR for a finite value of $r$. This is associated to the fact that the coupling $\la_s$ reaches the pole of the beta function at $\la_s=\la_0$. 

  \subsection{$\beta_s =$ 2-loops $\beta$ with $f_\la(\la_s) = \la_s$ and $f_A(A) = A$  }
   In this second case one still has $\la_s = \la$, $\log\mu = A$ and hence $\beta_s (\la_s) \equiv \beta(\la_s)$ but now $\beta$ is taken to be the two loops beta function. The solutions of the \eq{eom} for the relevant quantities are presented in Fig.~\ref{fig:backgrounds}. The curves for this setup are the dashed red ones.  In the top-left panel it is plotted the evolution of the coupling as a function of the energy scale ($\log \mu \sim - \log r$).
The geometry of the solutions (top-right and bottom-left panels) has the following qualitative behavior:  The geometry is still almost exactly $AdS$ but the space   ends in the IR for a finite value of $r$ due to the fast increase of the string coupling constant.

    \subsection{All-orders $\beta$ with modified $f_\la(\la_s)$ and $f_A(A) $  } \label{IRmodsec}
It is a simple matter to see that when using the beta functions described above in the right-hand side of  \eq{assol} one cannot derive \eq{Aas} if $f_\la$ and $f_A$ are identity functions. Another way to see why there is such a discrepancy is that, by taking  $f_\la$ and $f_A$ as identity functions, and upon integrating the beta functions in the right-hand side of \eq{assol} the derived geometry is such that the IR is reached for a finite value of the fifth coordinate $r$.  According to \cite{Gursoy:2007er} one obtains the particle spectrum  $m^2 \sim n^2$, for sufficiently large integer number $n$, which is analogous to the one arising in a one-dimensional quantum mechanical particle in a box. 

Hence, we modify the transition functions in the infrared together with the all-orders ansatz for the gauge theory YM beta function. In order to satisfy \eq{Aas} we require:
\be \label{TFbc}
  \la = f_{\la}(\la_s)\ \substack{\phantom{A}\\ \longrightarrow\\ \la_s \to +\infty}\ \la_0\ ; \qquad \log\mu= f_A(A)\ \substack{\phantom{A}\\ \longrightarrow\\ A \to -\infty} \log\Lambda \ . 
\ee
where $\Lambda$ is the energy where the evolution of the coupling constant, dictated by the beta function above, terminates as $\la \to \la_0$.
A simple choice for an explicit for of the transition function $f_\la$ is
\be \label{fldef}
 \inv{\la_s} = \inv{\la} + \inv{\la_0}\log\la + \mathrm{const.} \ ,
\ee
It is interesting to note that this transition function is, naively, the one that transforms the all-orders beta function in the one loop one if the $\lambda_s$ is interpreted as the gauge coupling in the one-loop scheme. This is exactly the transformation one often encounters in supersymmetric gauge theories. The one-loop function can be directly identified with the string beta function $\beta_s$ only if $f_A$ is the unity transition function. However, as we shall see momentarily, this is not case.

Having found $f_\la$ we are left with $f_A$ to determine via \eq{assol}, i.e.
\be
 A(\la_s) = f_A^{-1}\left(\inv{b_0 \la_s}+\log\Lambda\right) \ ,\label{realassol}
\ee
where we used  \eq{TFbc} to determine the integration constant. Analytic expressions can be derived in the deep IR and UV. In the IR regime using \eq{Aas} for the left-hand side of \eq{realassol} we obtain: 
\be
 f_A(A)\! - \! \log \Lambda \ \propto \  (-A)^{-3/8} e^{-3A/2} 
\ee
as $A \to -\infty$. In the UV we simply require that the function approaches the identity one, \ie, 
\be
 f_A(A) = A\left[1+ \morder{\frac{1}{A}}\right] \ ; \qquad A \to +\infty \ .
\ee
A simple example of a transition function $f_A$ that meets both the IR and UV requirements is
\be \label{fAdef}
 f_A(A) = \log \Lambda + \frac{2}{3}\log\left[1+(A^2+A_0^2)^{-3/16} e^{3A/2}\right] \ .
\ee
Here $A_0$ is a nonzero constant.

\begin{figure}[h]
\begin{center}
\flushleft \includegraphics[height=5.5 cm,width=7 cm]{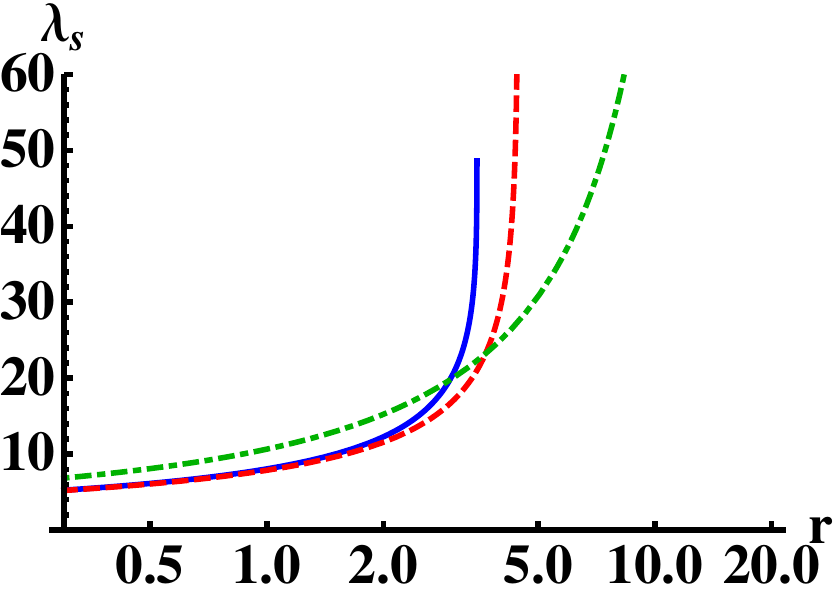}\hspace{2.cm}\includegraphics[height=5.5 cm,width=7 cm]{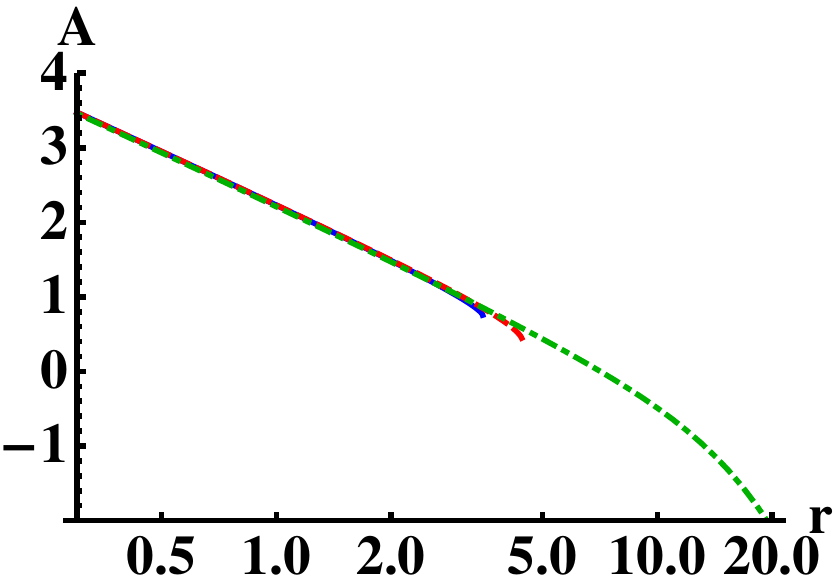} \\~\\
\includegraphics[height=5.5 cm,width=7 cm]{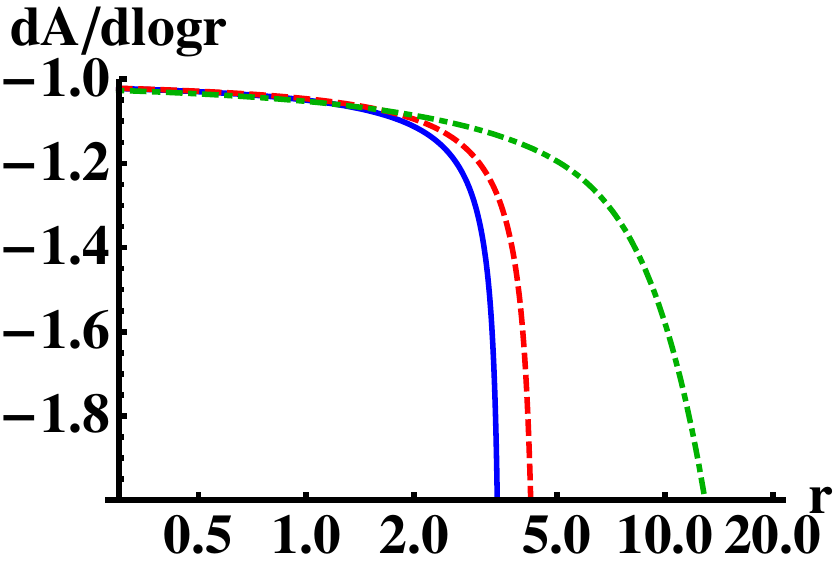}
\end{center}
\caption{The coupling constant lambda (top-left), the warp factor $A$ (top-left), and the logarithmic derivative of the warp factor (bottom-left) as a function of the conformal coordinate $r$. The curves are the solution for the three scenarios: the all orders beta function (blue continuous curve), the 't~Hooft beta function (dashed red curve), and the all orders beta function with IR matching (green dash-dotted curve). }\label{fig:backgrounds}\end{figure}

Having determined the transition functions we move to solve the equation of motions recalling that: 
\begin{equation}
\beta_s(\la_s) = \frac{\beta(\la)}{f^{\prime}_\la (\la_s)} f^{\prime}_A (A) = \beta_{\rm 1-loop}(\la_s) \frac{df_A}{dA} \ .
\end{equation}
  The solutions of the \eq{eom} for the relevant quantities are presented in Fig.~\ref{fig:backgrounds}. The curves for this case are the green dash-dotted.  In the top-left panel it is plotted the evolution of the coupling as a function of the energy scale ($\log \mu \sim - \log r$).
 Differently from the above solutions with identity transition functions, the warp factor (top-right and bottom-left panels) shrinks very rapidly towards the IR which is reached for infinite $r$. In the UV ($r \ll 1 $) the coupling in this scenario slightly deviates from the other two due discussed above due to the modification of the transition functions. 

We have also checked what happens to the glueball spectra in the various scenarios. Even the lowest states turn out to be sensitive to the shape of the background in the IR and hence to the details of the transition functions.

\section{Adding fermions: The Holographic Conformal Window}

It would be interesting to include matter in the five dimensional setup. Usually one adds a small number ($N_f \ll N$) of fermions in the fundamental representation of the gauge group by introducing pairs of probe flavor branes that are assumed not to affect the background (see the discussion in \cite{Gursoy:2007er} for adding fermions in the present background). The fermion dynamics is described by Dirac-Born-Infeld and Chern-Simons actions on a fixed background. However, typically $N_f/N$ is not small for interesting theories, and one should consider the backreaction on the metric due to the flavor branes, which quickly becomes rather involved. Adjoint fermions could possibly be added by introducing extra bulk dimensions, in close analogy to the original idea of Maldacena \cite{Maldacena:1997re}.

\begin{figure}[h]
\begin{center}
\flushleft \includegraphics[height=5.5 cm,width=7 cm]{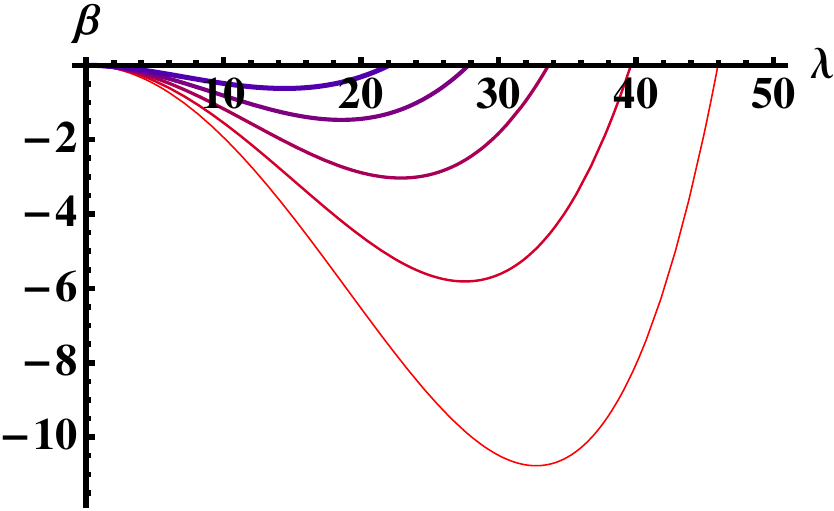}\hspace{2.cm}\includegraphics[height=5.5 cm,width=7 cm]{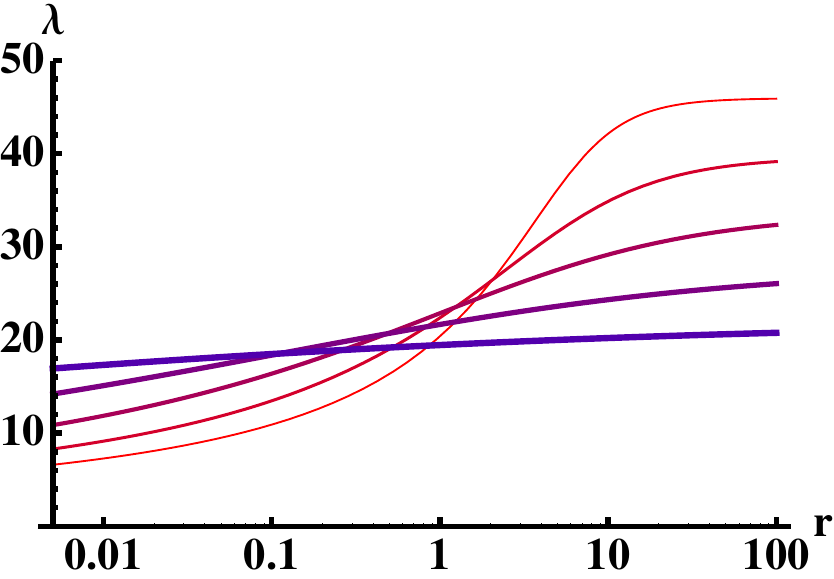} \\~\\
\includegraphics[height=5.5 cm,width=7 cm]{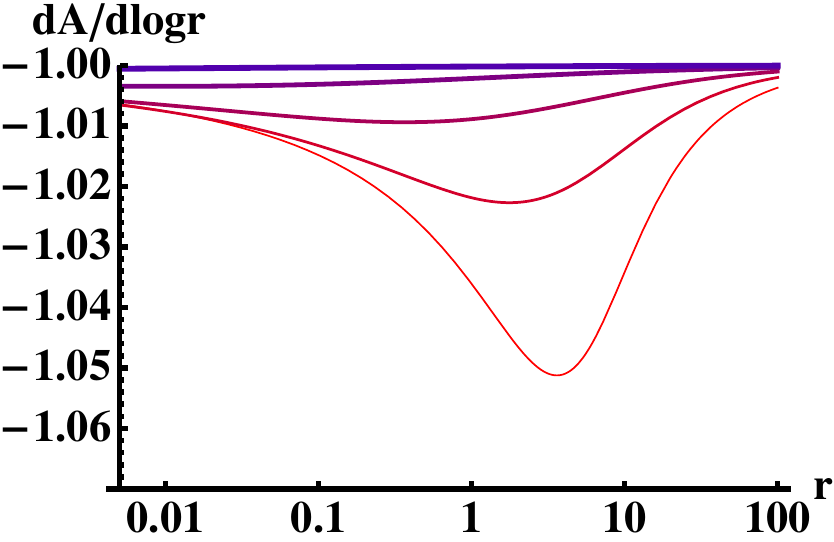}
\end{center}
\caption{The beta function (top-left), the coupling constant $\la$ (top-right), and the logarithmic derivative of the warp factor (bottom-left) as a function of the conformal coordinate $r$ for $N_f$ flavor QCD with $N_f$ assuming the values from $9$ (red thin curves) to $13$ (blue thick curves). }\label{fig:fermions}\end{figure}

In this paper we will not try to solve the seemingly difficult question of adding matter in this setup, but restrict to a very simple discussion. We shall not add any new dynamical degrees of freedom but only consider the effect of matter on the dilationic-induced field as predicted by the conjectured all-orders beta function.  The hope is that this method should correctly capture the qualitative influence of the matter fields on the geometry of the dual five dimensional setup. 

Recall that in the presence of matter, the all orders beta function depends on the anomalous dimension of the fermion mass $\gamma(g)$ which is unknown at strong coupling. However, for our purposes it is sufficient to adopt the  two-loop result in \eq{gamma}. 

We recall that extra information is needed to determine the exact boundary of the conformal window \cite{Ryttov:2007cx}, i.e. the lowest number of flavors below which chiral symmetry must break. This is translated in the largest possible value the anomalous dimension can have at the IRFP after which the theory breaks conformality. We have already discussed that $\gamma$ cannot, in any case, exceed the numerical value of two. This sets the lowest bound for $SU(N)$  gauge theories with fermions in any representation $r$ to be $N_f=11N/(8T(r))$ which corresponds to $33/4 = 8.25$ for $N=3$ and fundamental representation. It may very well be that the maximum value for the anomalous dimension is $\gamma=1$  (see  \cite{Sannino:2009za,Sannino:2008ha} for the reasons behind this choice) and in this case the critical number of flavors is $N_f = 11N/(6T(r))$ corresponding to $11$ for $N=3$ and fermions in the fundamental representation.  Alternatively one can use the exact solutions of the dual gauge theories to gain insight on this problem \cite{Sannino:2009qc,Sannino:2009me}.  We will restrict here to the number of flavors within the largest possible extent of the conformal window.

\begin{figure}[h]
\begin{center}
\flushleft \includegraphics[height=5.5 cm,width=7 cm]{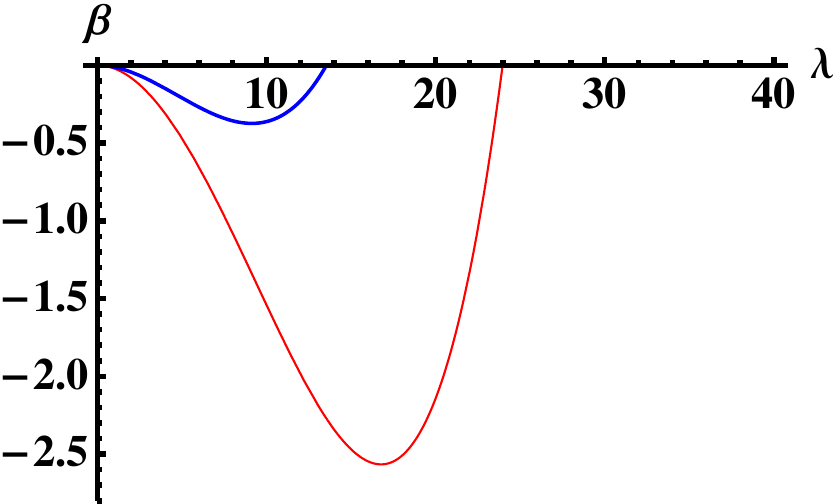}\hspace{2.cm}\includegraphics[height=5.5 cm,width=7 cm]{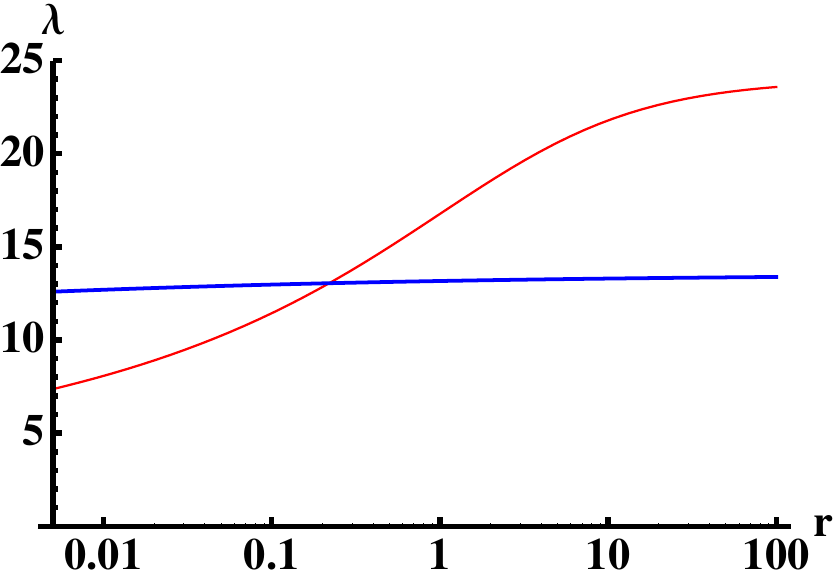} \\~\\
\includegraphics[height=5.5 cm,width=7 cm]{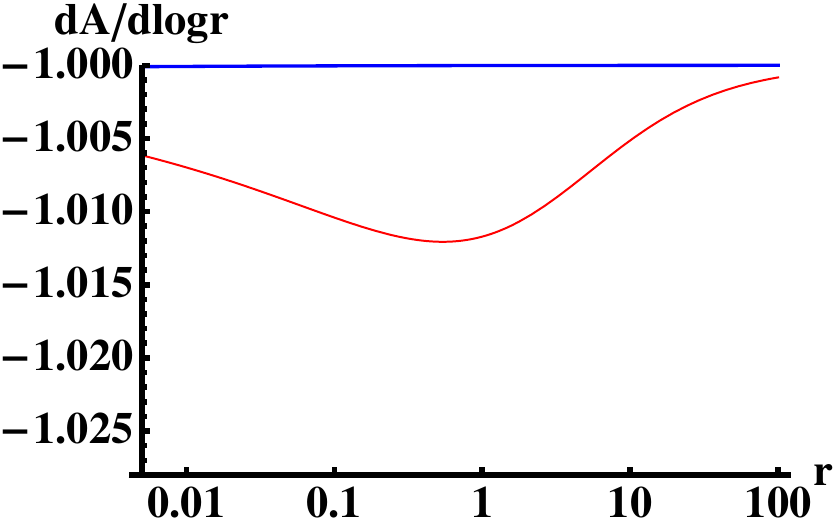}
\end{center}
\caption{The same as Fig.~\ref{fig:fermions} but for quarks in the adjoint (or symmetric) representation of $SU(2)$. $N_f$ assumes the value $1.5$ (red thin curves) or two (blue thick curves).}
\label{fig:fermionsadj}
\end{figure}

As first example we calculate the geometry for an $SU(3)$ gauge theory with $N_f$ Dirac fermions in the fundamental representation in the following range for flavors: $ 8.25 < N_f<16.5$, where the upper end of the conformal window coincides with the loss of asymptotic freedom. {}For this initial exploration we set the transition functions to the identity. This seems to be a reasonable approximation given that the space will result to be of $AdS$ form both in the UV and  in the deep IR. Using the expression above for the two-loop anomalous dimension, the explicit form of the beta function is: 
\be
 \beta_{\rm QCD} (\la) = - \frac{\la^2}{8\pi^2} \frac{\frac{11}{3}-\frac{2 N_f}{9}-\frac{N_f \lambda
   }{54 \pi ^2}+\frac{N_f (10 N_f-303) \lambda ^2}{46656 \pi ^4}}{1-\frac{(51-5 N_f) \lambda }{8 \pi ^2 (33-2 N_f)}} \ .
\ee
The UV boundary conditions, as explained in the appendix, are set by expanding the beta function to two-loops whose universal coefficients are: 
\be
 b_0 =  \frac{33-2N_f}{72\pi^2} \ ; \quad b = \frac{57 N_f-459}{(33-2 N_f)^2}  \ .
\ee
The beta function and the geometry are plotted in figure~\ref{fig:fermions} for $N_f=9 \ldots 13$. We discover that the latter always approaches an $AdS$ behavior, both in the deep IR and in the UV. This is most likely a generic feature of any background geometry dual of a gauge theory with an IRFP. We repeated the analysis for the case in which we have  $N_f$ Dirac  fermions transforming according to the adjoint representation of the $SU(2)$ gauge group. These are the {\it Minimal Walking Technicolor} models extensively studied via lattice simulations \cite{Catterall:2007yx,Catterall:2008qk,DelDebbio:2008zf,Hietanen:2008vc,Hietanen:2009az,Pica:2009hc,Catterall:2009sb} \footnote{Searches for the conformal
window in theories with fundamental representation
quarks have also received recent attention 
\cite{Appelquist:2009ty,Appelquist:2009ka,Fodor:2009wk,Fodor:2008hn,
Deuzeman:2009mh}}. The maximum extent of the conformal window here is $ 1.375<N_f<2.75 $. We present the plots for the running and the background geometry for $N_f = 1.5$ and $2$ in figure~\ref{fig:fermionsadj} .

\section{Conclusion}
We introduced an oversimplified five-dimensional framework which is meant to model the background geometry associated to ordinary YM as well as nonsupersymmetric gauge theories possessing an infrared fixed point with fermions in various representations of the underlying gauge group. The model is entirely based on the improved holographic approach, inspired by string theory, and on the recently conjectured all-orders beta function for any gauge theory with fermionic matter in any representation of the underlying gauge group. In the YM case the knowledge of the all-oders beta function is used to gain insight on the transition functions which are used to connect the renormalization scale and gauge coupling  with, respectively, the warp factor $A$ and the dilaton in the gravitational dual. We see that it is possible to determine the form of these transition functions in a way that one recovers the expected gauge theory properties from the gravitational dual. We then turn our attention to the very important problem of constructing a gravitational dual encoding the gauge theory number of flavors dependence. Here we used the closed form expression of the gauge theory beta function to transfer this information directly on the string-inspired one. This can, at best, be considered a phenomenological bottom up approach, however we find that it captures reasonably well the theoretical expectations. More specifically, we observe that the geometry becomes $AdS$ when approaching the ultraviolet and infrared regimes for the gauge theories in the conformal window. Besides, as the number of flavors increases, within the conformal window, we show that the geometry becomes more and more of $AdS$ type over the entire energy range. 

This setup can be readily generalized to account for a number of interesting applications ranging from high temperature investigations of the quark-gluon plasma in gauge theories with IRFP to the investigations of the effects of introducing a mass term in the beta function \cite{Dietrich:2009ns} as well modification of our conjectured beta function \cite{Antipin:2009wr}. 

\acknowledgements
M.J. and F.S. have been partially supported by the Marie Curie Excellence Grant under contract MEXT-CT-2004-013510.  M.J. has also been supported by the Villum Kann Rasmussen foundation. We thank for discussions J. Bechi, D.D. Dietrich, R. Foadi, M.T. Frandsen, P. Hoyer, C. Hoyos, N. Jokela, E. Kiritsis, C. Kouvaris,  F. Nitti, S. Nowling, T.A. Ryttov and T. Tahkokallio. 
 
\appendix
\section{On the solutions of the differential equations}
 The solutions for the equations of motion \eq{eom} involve, in general, three integration constants. These are  a shift $\bar{A}$ in the warp factor $A$ and two parameters $\Lambda$ and $\delta r$ associated to a linear transformation of $r$. Denoting with $A_*(r)$ and $\la_*(r)$ the particular solutions of \eq{eom}, the general solution is 
\bea \label{gensol}
 A(r)   &=& \bar A + A_*\left(\Lambda(r-\delta r)\right) \ ,\nn\\
 \la_s(r) &=& \la_*\left(\Lambda(r-\delta r)\right) \ .
\eea 
Here the free parameters $\bar A$, $\Lambda$, and $\delta r$ can be related to (the logarithm of) the $AdS$ radius $\log\ell$,  (the inverse) of the units of $r$ or the QCD energy scale, and the position of the UV singularity, respectively (see \cite{Gursoy:2007er} for a thorough discussion). 
In the UV the above general result is reflected in the asymptotic expansions, which can be written as
\bea \label{UVseries}
 A(r) &=&  -\log \frac{r}{\ell}  + \frac{4}{9} \inv{\log r \Lambda} - \frac{4}{9} b \frac{\log(-\log r \Lambda)}{(\log r \Lambda)^2} + \frac{2+4 b - 4 K}{9}\inv{(\log r \Lambda)^2} + \morder{\inv{(\log r \Lambda)^3}} \nn\\
 b_0 \la_s(r) &=& - \inv{\log r \Lambda} + b \frac{\log(-\log r \Lambda)}{(\log r \Lambda)^2} + \frac{K}{(\log r \Lambda)^2} + \morder{\inv{(\log r \Lambda)^3}} \ .
\eea 
Here we fixed $\delta r$ by requiring that the singularity is located at $r=0$. The parameter $K$ is not indepedent but can be eliminated by rescaling $\Lambda$. Therefore we can set it to zero. The constant $b$ is proportional to the next-to-leading coefficient in the string beta function,
\be
 \beta_s(\la_s) = -b_0 \la_s^2 + b\ b_0^2 \la_s^3 + \cdots
\ee
The analysis above modifies if we the transition functions are not the identity. For example the modification in the UV amounts, in general, to a change in the second coefficient of the string beta function which otherwise is simply related to the  two-loop coefficient of the gauge beta function which for YM is
\be 
 b = - \inv{b_0 \la_0} = -\frac{51}{121} \simeq -0.42 \ .
\ee
In the main text we have investigated the case in which the transition functions are not unity and there one finds $b=1/4$. The parameter $\Lambda$ fixes the scale of $r$ and hence controls the convergence radius of the asymptotic expansions. We take $\Lambda=0.1$ which is seen to produce deviation from the UV behavior around $r \sim 1$.  We have also tuned $\ell$  to obtain the desired asymptotics in the IR when the transition functions are nontrivial. 

We match the numerical solutions to the asymptotic series \eq{UVseries} at $r=10^{-5}$ \footnote{As a consistency check, this reproduces the UV singularity at $r=0$ within the accuracy of $~10^{-9}$.}. Taking $A_0=1$ in \eq{fAdef} the IR asymptotics is correct in the scenario of section~\ref{IRmodsec} for $\ell \Lambda \simeq 1.064$. The solutions for the YM analysis are plotted in figure~\ref{fig:backgrounds}. In addition to the coupling constant $\la_s$ and the warp factor $A$ we show the behavior of the logarithmic derivative $dA/d\log r$ where the deviation from the $AdS$ metric (which obeys $dA/d\log r = -1$) becomes clearly visible.

\end{document}